\documentclass[aps,pra,preprint]{revtex4-2}

\usepackage{graphicx}
\usepackage{xcolor}
\usepackage{amsmath}
\usepackage{physics}
\usepackage{hyperref}

\newcommand{\omegar}{\omega_\mathrm{r}}
\newcommand{\oeff}{\Omega_\mathrm{eff}}

\newcommand{\nbragg}{n_\mathrm{Bragg}}
\newcommand{\eq}[1]{Eq.(\ref{eq:#1})}
\newcommand{\fig}[1]{Fig.(\ref{fig:#1})}
\newcommand{\unit}[1]{\,\mathrm{#1}}

\begin{document}


\title{Serrodyne Matterwave Optics}


\author{Jack Roth}
\email{jack\_roth@berkeley.edu}
\author{Madeline Bernstein}
\author{Andrew Christensen}
\author{Yuno Iwasaki}
\author{Holger Mueller}
\affiliation{Department of Physics, University of California, Berkeley}

\author{Zaire Sprowal}
\author{Robert Compton}
\affiliation{Safran Federal Systems}


\date{\today}
\begin{abstract}
Bragg diffraction for atom interferometry conventionally requires two laser frequencies whose difference makes the two-photon transition resonant. We show that this frequency difference can instead be generated by serrodyne modulation, allowing Bragg pulses to be driven with a single-frequency laser. The serrodyne modulation is performed using an acousto-optic modulator driven with a phase-modulated RF drive, and generated kilohertz-scale frequency shifts on top of the AOM carrier frequency. While in general serrodyne modulation generates strong unwanted frequency components, we find conditions so that the Bragg diffraction pulses, and atom interferometers constructed from them, are robust to these imperfections, opening the door to the development of ultra compact, single frequency Bragg diffraction based atom interferometers with serrodyne modulation implemented by a position modulated retroreflection mirror.
\end{abstract}


\maketitle

\section{Introduction}

Atom interferometers are used in a variety of applications, from testing fundamental physics \cite{Tino2021,Morel2020,Overstreet2022} to field measurements of the local acceleration of gravity \cite{Lyu2022,Wu2019,Devani2020} to inertial sensing of acceleration and rotation \cite{Gustavson2000,Dubetsky2006,Saywell2023,Salducci2024}. To make measurements, atom interferometers split free particles into superpositions of spatially separated wavepackets that accumulate a phase difference proportional to a quantity of interest and then interfere. In light-grating atom interferometers, the needed matterwave optical elements (beamsplitters and mirrors) are formed by counterpropagating laser beams: in Raman \cite{Kasevich1991} or Bragg \cite{Peik1997,Mueller2008,Clad2009,Kovachy2012,Pagel2020,Bguin2022} transitions, atoms absorb a photon from one beam with impulse $\hbar k$ and then emit a photon into the other beam with an impulse of $\hbar k$. Because the two beams are counterpropagating the atom undergoes a net momentum shift of $2\hbar k$. The beams are often formed by retroreflecting a two-frequency laser beam \cite{Sugarbaker2013,Tennstedtnavi,Rosi2014,Mazzoni2015}, leading to a stable phase between each counterpropagating frequency pair in the frame of the retroreflection mirror. To satisfy energy conservation, the frequency difference between the two beams must be $\delta=\pm4\omegar$ (plus the hyperfine splitting, in the case of Raman transitions), where $\omegar=\hbar k^2/2m$. The sign of $\delta$ determines the direction of the kick. Unfortunately, however, the symmetry of this approach means that atoms at rest will be equally likely to receive an impulse into either direction \cite{Lvque2009,Giese2013,Jenewein2022,Hartmann2020}, unless the symmetry is broken by another mechanism, like free fall \cite{Gauguet2008,parker2018}. This zero-velocity degeneracy is especially problematic for operation in space \cite{Sorrentino2010,Frye2021,Chiow2017}, multi-axis sensors \cite{LeDesma2025,Rakholia2014,Stolzenberg2025,LeDesma2025_2}, and compact quantum sensors \cite{Wu2019,Lee2022}. Previous work has overcome this degeneracy by polarization engineering of Raman transitions \cite{Bernard2022}, or by using broadband (chirped \cite{Perrin2019} or mode-locked \cite{Solaro2022}) lasers where the propagation delay generates an asymmetry. These solutions come at the cost of increasing laser system complexity, or do not broadly work for both Bragg and Raman transitions.

Serrodyne-modulation uses a high-frequency sawtooth waveform to phase modulate a laser, producing an effective frequency shift \cite{Cumming1957,Houtz2009,Johnson2010}. Implementing serrodyne modulation by modulating the position of the retroreflection mirror could, in principle, be used to break the zero-velocity degeneracy. However, achieving sufficiently fast modulation is not possible because of the inertia of a high-quality mirror, which results in a loss of fidelity in the beamsplitter process. However, controlling the phase of the sawtooth waveform used for serrodyne-modulation and the start of the beamsplitter pulse, the dynamics of the atom’s state under the influence of the serrodyne-modulated light allows for high-fidelity Bragg diffraction even with the seemingly poor-quality modulation that can be achieved with existing mirror agility. Here, we present a Mach-Zehnder atom interferometer composed of serrodyne matterwave optics which break the zero-velocity degeneracy through serrodyne-modulation of the retroreflection mirror. We implement the Mach-Zehnder interferometer experimentally in an atomic fountain, where the effect of the mirror motion is simulated by an optical modulator. Our experiment is subject to (and therefore allows studying of) the same beam size, cloud temperature, and cloud size limitations found in the moving mirror system, and allows varying the quality of the serrodyne-modulation without the limitation of mirror bandwidth. We show that our experiment is described by the same equations of motion a stationary atom cloud and a moving mirror are subject to. We experimentally test the limits on fall time and amplitude that the serrodyne-modulation sawtooth waveform must adhere to, and measure the performance of a Mach-Zehnder interferometer composed of serrodyne-modulated Bragg beamsplitters. The method is laser-power efficient and optically simple, requiring only one laser frequency. This is advantageous for field-based atom interferometers, as it trades laser complexity for a moving mirror. This method opens the door to creating compact atom interferometers with a single laser for 3D sensing and for space applications.

\section{Serrodyne-modulation equations of motion}
\label{sec:serrodyne_optics}

\begin{figure}
    \centering
    \includegraphics[width=\linewidth]{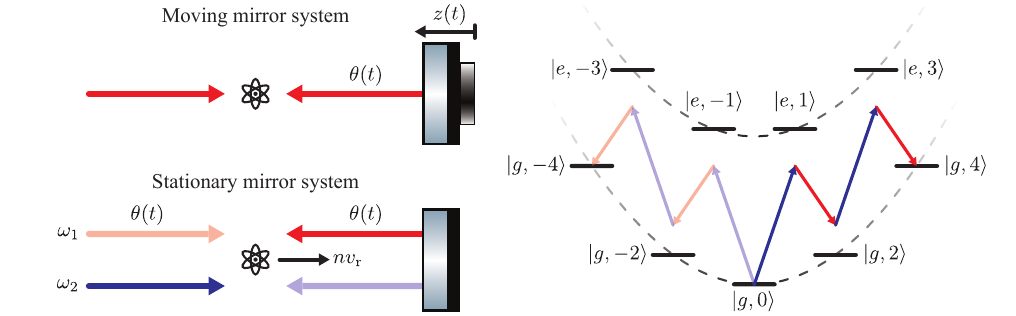}
    \caption{\textit{Left:} Two experiment configurations are shown: the moving mirror system and the stationary mirror system. In the moving mirror system, a single-frequency laser is retroreflected off of a moving mirror. The retroreflected field has the same frequency as the incident field, and a relative phase of $2kz(t)$. This forms a moving standing wave in the lab frame. In the stationary mirror system, a two-frequency laser is retroreflected off of a stationary mirror and is incident on a moving atom. The atom is moving fast enough such that the incident $\omega_1$ and reflected $\omega_2$ are Doppler shifted enough to suppress any multiphoton process. This means that both configurations have an incident beam without phase modulation, and a retroreflected beam with phase modulation $\theta(t)$. By applying a sawtooth waveform to $\theta(t)$, we drive a Bragg process by effectively shifting the frequency of the retroreflected beam to make the Bragg process resonant. \textit{Right:} The Bragg diffraction level structure. Initially stationary atoms in the ground state are coupled via second-order Bragg diffraction to the state with momentum $4\hbar k$. With serrodyne-modulated Bragg, the red field is made resonant using phase modulation.}
    \label{fig:system_diagram}
\end{figure}

\begin{figure}
    \centering
    \includegraphics[width=0.5\linewidth]{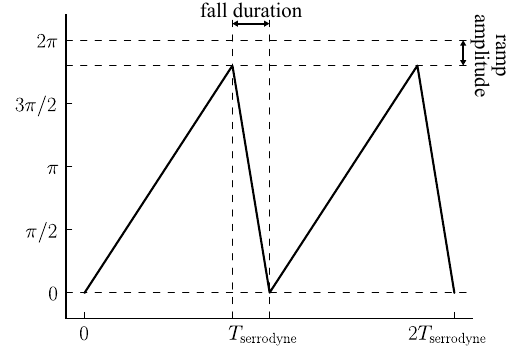}
    \caption{The sawtooth waveform $\theta(t)$ used for serrodyne-modulation is characterized by several figures of merit: the ramp amplitude (ideally $2\pi$), the sawtooth period $T_\mathrm{serrodyne}$ (which along with the ramp amplitude sets the effective frequency shift), and the fall duration (ideally zero). Here we show a sawtooth waveform with an incorrect ramp amplitude and a non-zero fall duration.}
    \label{fig:serrodyne_error_sources}
\end{figure}

Typically the frequency difference $\delta$ between the counterpropagating lasers is generated using an acousto-optic modulator (AOM). However, in the retroreflection scheme it is produced by modulating the retroreflection mirror position in time to imprint a time-dependent phase of $\delta t$ on the retroreflected beam as shown in \fig{system_diagram}. The mirror moves at a constant velocity to produce a phase $\theta(t)$ on the retroreflected beam. Once a phase of $2\pi$ has been reached, the mirror can be quickly moved back to its initial position. This cycle can be repeated to produce a frequency difference of $\delta$ between the two beams. The mirror must be reset to its initial position fast enough to avoid disturbing the Bragg diffraction dynamics. The phase produced by the moving mirror as a function of time is shown in \fig{serrodyne_error_sources}.

We consider the equations of motion for a two-level system in a far-detuned single-frequency laser retroreflected off of a moving mirror. The retroreflected field has phase $\theta(t)=2kz(t)$, where $z(t)$ is the position of the mirror as a function of time. The equations of motion for the ground state of this system are:
\begin{eqnarray}
    \dot{c}_n(t)=&-i4n^2\omegar c_n(t)-i2\oeff c_n(t) \nonumber \\
    &-i\oeff\exp(-i\theta(t))c_{n-1}(t) \nonumber \\
    &-i\oeff\exp(i\theta(t))c_{n+1}(t) \label{eq:eom_mirror_main}
\end{eqnarray}
where $\omega_r=\hbar k^2/2m$ is the recoil frequency, $k$ is the laser wavenumber, $m$ is the atomic mass, $\oeff=\Omega^2/4\Delta$ is the effective Rabi frequency, $\Delta$ is the single photon detuning, $\Omega$ is the single-photon Rabi frequency produced by each beam, $n$ indexes the momentum state in units of $2\hbar k$, and $c_n$ are the momentum state amplitudes.

A perfect sawtooth waveform with zero fall duration would modulate the mirror position with $z(t)=(\lambda t/2T_\mathrm{serrodyne})\,\mathrm{mod}\,(\lambda/2)$ and corresponding phase $\theta(t)=(2\pi t/T_\mathrm{serrodyne})\,\mathrm{mod}\,(2\pi)$. Because the exponential terms are periodic in $2\pi$ we can drop the modulus and write
\begin{eqnarray}
    \dot{c}_n(t)=&-i4n^2\omegar c_n(t)-i2\oeff c_n(t) \nonumber \\
    &-i\oeff\exp(-i\delta t)c_{n-1}(t) \nonumber \\
    &-i\oeff\exp(i\delta t)c_{n+1}(t)
\end{eqnarray}
where $\delta=2\pi/T_\mathrm{serrodyne}$. These are the usual equations of motion describing Bragg diffraction \cite{Mueller2008,Kovachy2010}.

The Bragg resonance condition is found by finding the value of $\delta$ that satisfies both energy and momentum conservation. For an initially stationary atom this corresponds to $\delta=4\nbragg\omegar$. We compute the required mirror motion for an initially stationary atom via
\begin{equation}
    z(t)=(\nbragg\omegar\lambda t/\pi)\,\mathrm{mod}\,(\lambda/2)
\end{equation}
In practice the mirror moves with some non-zero fall duration from $z=\lambda/2$ to $z=0$ as shown in \fig{serrodyne_error_sources}. During this non-zero fall the Bragg drive effectively pauses as the lattice is moving in the wrong direction. The drive resumes coherently when the ramp rises again. We find that fall durations comparable to the rise duration can still produce working beamsplitters, meaning that serrodyne-modulated Bragg pulses display useful behavior even when they poorly approximate normal Bragg diffraction. 

We note that there is a special case where the sawtooth period $T_\mathrm{serrodyne}$ is greater than the Bragg pulse width. In this regime the mirror creates an uninterrupted frequency shift by sweeping at a sustained rate of $\nbragg\omegar\lambda/\pi$.

A second special case occurs when the phase is modulated according to a symmetric triangle wave and the wave peak or trough occurs simultaneously with the peak of $\oeff(t)$. In this condition it is possible to create double-Bragg diffraction \cite{Lvque2009,Hartmann2020,Jenewein2022,Giese2013}, where an atom initially in state $\ket{0}$ is diffracted into $\ket{\nbragg}$ and $\ket{-\nbragg}$ with equal amplitudes. This beamsplitter configuration is of interest for other interferometer geometries.
 
\section{Serrodyne-modulation in an atomic fountain}

\begin{figure}
    \centering
    \includegraphics[width=0.8\linewidth]{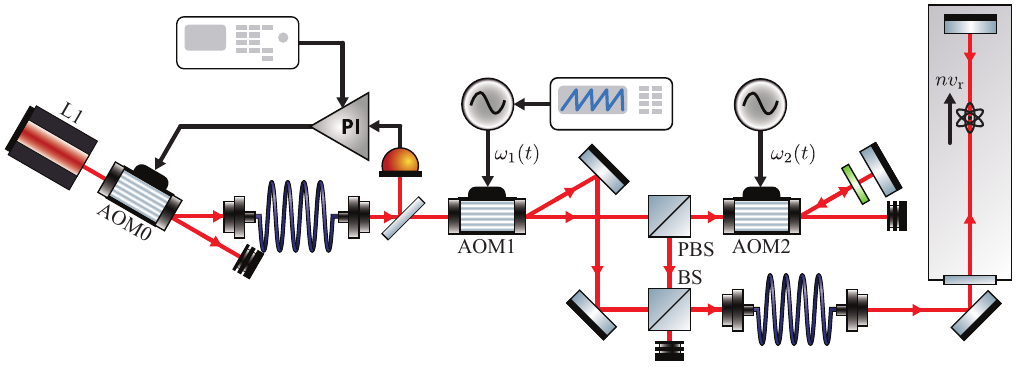}
    \caption{Optical chain used to perform phase modulation on the interferometry laser. A titanium:sapphire laser (L1) blue detuned $6675\unit{MHz}$ from the Cesium $6\,^2S_{1/2}\,F=3\to6\,^2P_{3/2}\,F=4$ transition is used to generate Gaussian pulses in time by amplitude modulating AOM0. The diffracted light passes through AOM1, which is driven with phase-modulated $\omega_1(t)$. The frequency $\omega_1(t)$ is generated by an RF synthesizer with a phase modulation port, which sets $\omega_1(t)=\omega_1+\partial\theta(t)/\partial t$, where $\omega_1=180.2274\unit{MHz}$ and $\theta(t)$ is set by a function generator programmed with a sawtooth waveform connected to the phase modulation port of the RF synthesizer. If normal Bragg diffraction is used instead of serrodyne-modulated Bragg diffraction, $\theta(t)=0$ and $\omega_1$ is set to whatever value brings the Bragg process into resonance. The diffracted order is then coupled into a fiber and sent to the experiment vacuum chamber. The undiffracted order goes through double-passed AOM2 driven by time-dependent frequency $\omega_2(t)$. The frequency $\omega_2(t)$ is set by a linearly swept DDS. The frequency sweep rate is set to cancel the Doppler shift that the launched atom cloud accrues due to gravitational acceleration. This ensures that in the frame of the atom cloud $\omega_1(t)-\omega_2(t)$ is constant. When serrodyne-modulation is used $\omega_2(t)=\omega_1$ in the frame of the atom cloud. The double-passed light is coupled into the same fiber and sent to the experiment vacuum chamber.}
    \label{fig:ifr_chain_diagram}
\end{figure}

To investigate tolerance to limitations of the mirror motion we perform serrodyne-modulated Bragg diffraction using a phase-modulated two-frequency retroreflected laser in an atomic fountain as displayed in \fig{system_diagram}. This system does not have a piezo-mounted mirror, but uses an AOM to reproduce the same form of phase modulation produced by the piezo-mounted mirror. This allows us to simulate a variety of sawtooth imperfections without being limited by the choice of piezo. In the appendix we show that the equations of motion for the stationary mirror system are equivalent to the equations of motion in \eq{eom_mirror_main}.

The experiment is performed in a cylindrical vacuum chamber $4\,\mathrm{m}$ long and $0.4\,\mathrm{m}$ in diameter. Cesium atoms are loaded into a 3D magneto-optical trap (MOT) from a room-temperature 2D MOT. After $1.15\unit{s}$ of loading from the 2D MOT, the 3D MOT coils are switched off, polarization gradient cooling is performed, and a molasses launch is used to launch the atoms directly upwards at $6.2\unit{m/s}$. $10\unit{ms}$ after launch the atoms enter a Raman sideband cooling region, where they are cooled for approximately $3\unit{ms}$ in a moving optical lattice (the lattice is at rest in the frame of the atoms). This is followed by an adiabatic transfer step which transfers the atoms from the $F=3,m_F=3$ hyperfine ground state to the $F=3,m_F=0$ state. Finally two consecutive Raman velocity selective pulses are used to select a vertical velocity distribution of atoms that is $0.1\times2\hbar k$ in width. The atoms are subjected to resonant light between each Raman pulse to blow away the untransfered atoms. At this stage there are approximately 50,000 atoms in the $F=3,m_F=0$ hyperfine ground state with a transverse velocity distribution of $7\unit{mm/s}$.

\begin{figure}
    \centering
    \includegraphics[width=0.5\linewidth]{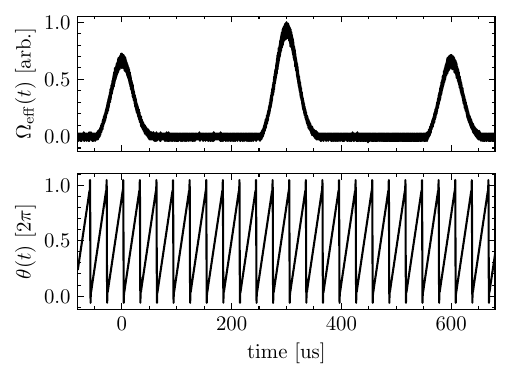}
    \caption{The time-dependent effective Rabi frequency $\Omega_\mathrm{eff}(t)$ (proportional to the pulse intensity) and the time-dependent phase $\theta(t)$ set by a sawtooth waveform recorded from a single interferometer shot. The sawtooth waveform rising ramp is configured for $\nbragg=4$. The $\Omega_\mathrm{eff}(t)$ displayed here was captured on a photodiode located behind a backpolished mirror directly before the vacuum chamber. The displayed $\theta(t)$ is the direct output of the function generator connected to the phase modulation port of the RF synthesizer that generates $\omega_1(t)$.}
    \label{fig:bragg_pulse_serr_waveform}
\end{figure}

The laser used to drive Bragg diffraction pulses propagates vertically and is aligned with the atom cloud in the transverse plane, and has waist $w_0=7.7\unit{mm}$ ($1/e^2$ radius). The atom cloud has a Gaussian extent of $\sigma=1.5\unit{mm}$ at the time Bragg diffraction is performed. The two-frequencies in the laser are prepared using the optical system described in \fig{ifr_chain_diagram}. The laser intensity (proportional to $\oeff(t)$) is modulated to produce a Gaussian intensity in time. The phase $\theta(t)$ of one of the laser fields is modulated with a sawtooth waveform in time to achieve serrodyne-modulation. \fig{bragg_pulse_serr_waveform} shows the phase modulation injected into the system along with the measured Bragg pulse intensity for a single experiment shot. In \fig{ifr_chain_diagram} three Bragg pulses are triggered to generate a Mach-Zehnder interferometer.

The first Bragg diffraction pulse is performed $225\unit{ms}$ after the molasses launch when the atoms are moving at $\sim4\unit{m/s}$. As the atoms are in different momentum states after Bragg diffraction pulses, they separate spatially. Roughly $1\unit{s}$ after the molasses launch the atoms fall through a resonant laser and their fluorescence is captured on a photodiode. Because each momentum state has spatially separated, multiple fluorescence peaks are detected corresponding to the momentum states output by the Bragg process. The fluorescence peak heights are used to compute the relative population in each momentum state.

\section{Serrodyne Bragg beamsplitters}
\label{sec:serrodyne_single_bragg}

Serrodyne-modulated single Bragg pulses are implemented in the experiment by driving $\theta(t)$ with a sawtooth waveform with a constant rising ramp rate of $4\nbragg\omegar$, a variable duration falling ramp, and a variable peak amplitude. We investigate the robustness of Bragg diffraction to variations in the falling ramp duration and peak amplitude. For each falling ramp duration and peak amplitude we measure the population in each output momentum state and compute the standard two-level population inversion $w=(|c_{\nbragg}|^2-|c_0|^2)/(|c_{\nbragg}|^2+|c_0|^2)$. The inversion captures the ability of the pulse to transfer population from the initial momentum state to the target momentum state. The inversion does not capture leakage to outside states.

We perform experiments for $\nbragg=2,3,4$. When the ramp amplitude is $2\pi$ exactly, the rising ramp duration of the sawtooth waveform is $60.5\unit{us}$, $40.3\unit{us}$, and $30.3\unit{us}$ for $\nbragg=2$, $3$, and $4$, respectively. The Bragg pulse intensity is given by a Gaussian profile in time with $\sigma=40\unit{us}$. Each Bragg pulse is therefore subject to multiple falling ramps of the sawtooth waveform.

\begin{figure}
    \centering
    \includegraphics[width=0.5\linewidth]{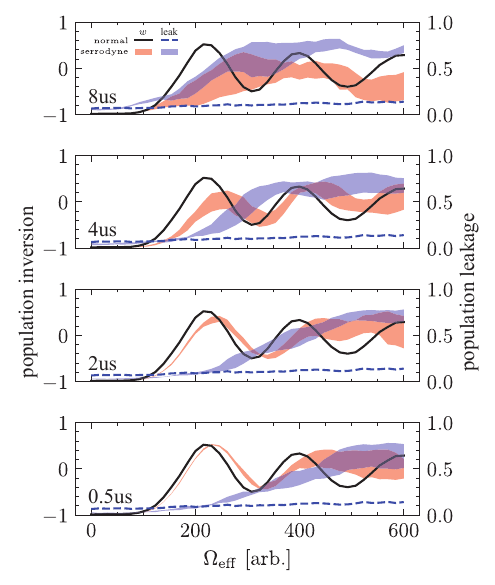}
    \caption{Measured population inversion for $\nbragg=3$ as a function of effective Rabi frequency for several sawtooth waveform fall durations. For each fall duration ten sawtooth waveform phase values between $0$ and $2\pi$ are recorded. The extent between the minimum and maximum value of the inversion and leakage are shaded in red and blue, respectively. The inversion and leakage of a normal non-serrodyne Bragg pulse are shown in solid black and dashed black, respectively. For $\nbragg=3$ the rising sawtooth waveform duration is $40\unit{us}$. As the fall duration gets longer the effect of the waveform phase on leakage and inversion becomes larger. For a $0.5\unit{us}$ fall duration the phase has little impact and the serrodyne-modulated Bragg pulse performs similarly to the normal Bragg pulse up to the $\pi$-pulse condition.}
    \label{fig:bragg_phase_scan}
\end{figure}

To investigate the effect of the sawtooth waveform phase relative to the Bragg pulse intensity we perform scans of the effective Rabi frequency at 10 phase shifts between $0$ and $2\pi$ for several falling ramp durations. The results of this scan are given in \fig{bragg_phase_scan}, where we shade a region set by the minimum and maximum observed inversions across all phases. From these scans it is clear that the initial waveform phase has a large effect on the single Bragg pulse dynamics, changing performance from close to a normal Bragg pulse to effectively not working.

\begin{figure}
    \centering
    \includegraphics[width=0.5\linewidth]{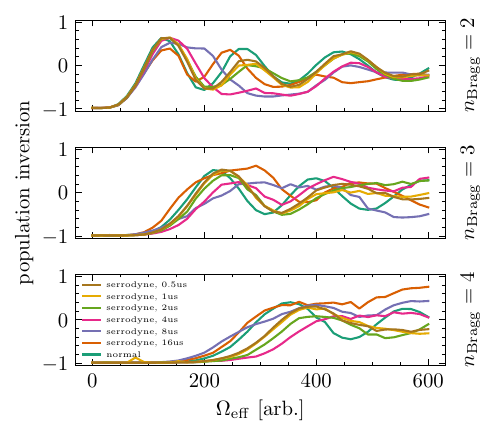}
    \caption{Measured population inversion as a function of effective Rabi frequency for several falling ramp durations and Bragg orders. The normal line denotes a Bragg diffraction pulse performed without serrodyne-modulation. The times in the legend denote the falling ramp duration.}
    \label{fig:bragg_fall_scan}
\end{figure}

To understand the performance as a function of fall duration we plot inversion as a function of $\oeff$ for the waveform phase that produces behavior closest to the normal (non-serrodyne) Bragg pulse. The ``closest'' phase is the one that minimizes the sum of squares for $\oeff$ between $0$ and the $\pi$ pulse condition for the normal Bragg pulse inversion curve and a serrodyne-modulated Bragg pulse inversion curve. Note the $\pi$ pulse condition is given by the first peak in the normal Bragg pulse inversion curve. This result is given in \fig{bragg_fall_scan} for several fall durations and Bragg orders. For $\nbragg=2$ it seems all fall durations are approximately equivalent between $0$ and the $\pi$ pulse condition. For $\nbragg=3$ we start to see substantial deviation from the normal $\pi$ pulse condition when the fall duration is greater than $2\unit{us}$, although all fall durations still reach a similar and usable maximum inversion. Similar conclusions are drawn for the $\nbragg=4$ case.

\begin{figure}
    \centering
    \includegraphics[width=0.5\linewidth]{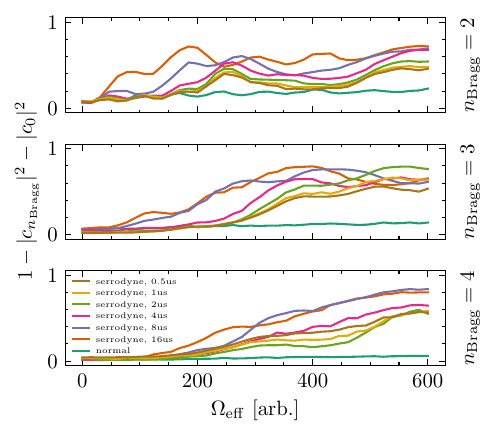}
    \caption{Measured population leakage to unwanted momentum states as a function of effective Rabi frequency for several falling ramp durations and Bragg orders. The normal line denotes a Bragg diffraction pulse performed without serrodyne-modulation. The times in the legend denote the falling ramp duration.}
    \label{fig:bragg_fall_leakage}
\end{figure}

We plot leakage to other states in \fig{bragg_fall_leakage}. We see that there is minimal leakage for the case of the normal Bragg pulse across all Bragg orders. Serrodyne-modulated pulses with fall durations greater than $2\unit{us}$ start to deviate from the normal Bragg pulse leakage before the $\pi$ pulse condition. If a longer fall duration is required than $2\unit{us}$, it will come at the cost of producing additional, and likely unwanted, momentum states at the interferometer output. However, as we show in section \ref{sec:serrodyne_mzi}, these fall durations can still produce coherent interferometers.

\begin{figure}
    \centering
    \includegraphics[width=0.5\linewidth]{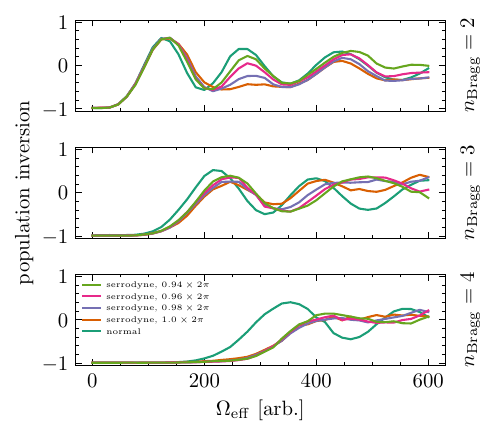}
    \caption{Measured population inversion as a function of effective Rabi frequency for several ramp amplitudes and Bragg orders. All serrodyne-modulated Bragg pulses have a falling ramp duration of $5\unit{us}$. The normal line denotes a Bragg diffraction pulse performed without serrodyne-modulation.}
    \label{fig:bragg_amp_scan}
\end{figure}

\begin{figure}
    \centering
    \includegraphics[width=0.5\linewidth]{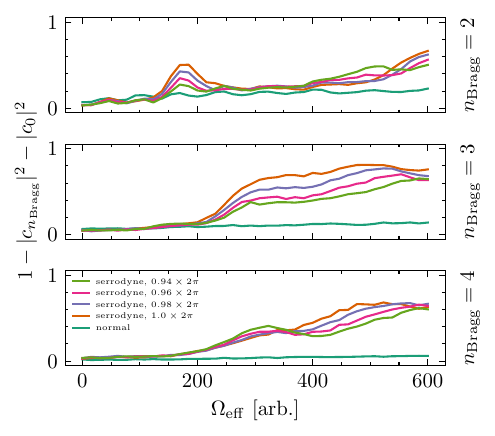}
    \caption{Measured population leakage to unwanted momentum states as a function of effective Rabi frequency for several ramp amplitudes and Bragg orders. All serrodyne-modulated Bragg pulses have a falling ramp duration of $5\unit{us}$. The normal line denotes a Bragg diffraction pulse performed without serrodyne-modulation.}
    \label{fig:bragg_amp_leakage}
\end{figure}

We now consider the performance of the serrodyne-modulated Bragg pulse as a function of the sawtooth waveform amplitude. \fig{bragg_amp_scan} shows the inversion as a function of effective Rabi frequency at a fall duration of $4\unit{us}$ for ramp amplitudes of $1\times2\pi$, $0.98\times2\pi$, $0.96\times2\pi$, and $0.94\times2\pi$, as well as for a normal Bragg pulse. For higher Bragg orders incorrect ramp amplitudes have a relatively small effect on the inversion. For lower Bragg orders incorrect ramp amplitudes appear to suppress the oscillation in the inversion as a function of $\oeff$. We plot leakage to other states in \fig{bragg_amp_leakage}, from which we see that lowering the ramp amplitude suppresses leakage to other states.

\section{Serrodyne Mach-Zehnder interferometer}
\label{sec:serrodyne_mzi}

\begin{figure}
    \centering
    \includegraphics[width=0.5\linewidth]{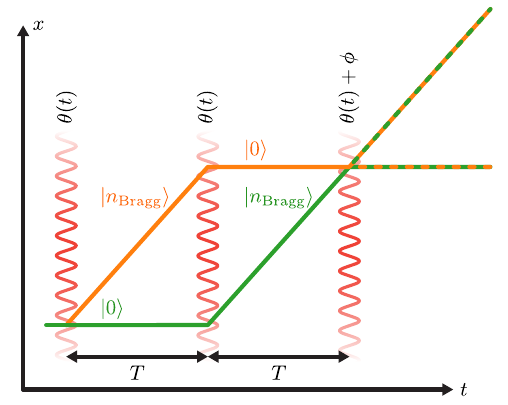}
    \caption{A spacetime diagram of the Mach-Zehnder interferometer (MZI) geometry. An atom described by a free-particle wavepacket enters the interferometer in momentum state $n=0$ (labeled via $\ket{0}$ in the diagram). The atom is subjected to a $\pi/2$--$\pi$--$\pi/2$ sequence of Bragg pulses at $t=0$, $t=T$, and $t=2T$. This sequence causes the wavepacket to split into a superposition of momentum states which each traverse different paths (either green or orange, as shown in the diagram). At the time of the final pulse the wavepackets interfere, yielding the populations described in the text. For a serrodyne-modulated Bragg pulse, $\theta(t)$ is given by a sawtooth waveform, while for a normal Bragg pulse $\theta(t)=0$. During the last Bragg pulse an additional phase modulation $\phi$ is applied to the field for both serrodyne-modulated and normal Bragg pulses. Scanning $\phi$ allows an interference fringe to be traced out.}
    \label{fig:mzi_geometry}
\end{figure}

To form a Mach-Zehnder interferometer (MZI) we perform three serrodyne-modulated Bragg diffraction pulses: a beamsplitter, a mirror, and a beamsplitter. The pulse spacing $T$ is set to an exact integer multiple of the sawtooth period to ensure each pulse experiences the same initial sawtooth waveform phase. The integer multiple is selected to make $T$ as close to $300\unit{us}$ as possible. The interferometer phase is scanned by introducing a phase shift $\phi$ to the final serrodyne-modulated Bragg pulse. The interferometer geometry is shown in \fig{mzi_geometry}. Due to small vibrations in the retroreflection mirror the interferometer phase varies shot-to-shot. This variation is averaged away by running multiple shots for each phase value $\phi$. The MZI can also be implemented using normal Bragg pulses. This is accomplished by shifting $\omega_1$ as described in \fig{ifr_chain_diagram}. Note that the sawtooth waveform phase $\theta_\mathrm{serrodyne}$ is distinct from the phase $\phi$ applied to the last pulse of the interferometer.

\begin{figure}
    \centering
    \includegraphics[width=0.5\linewidth]{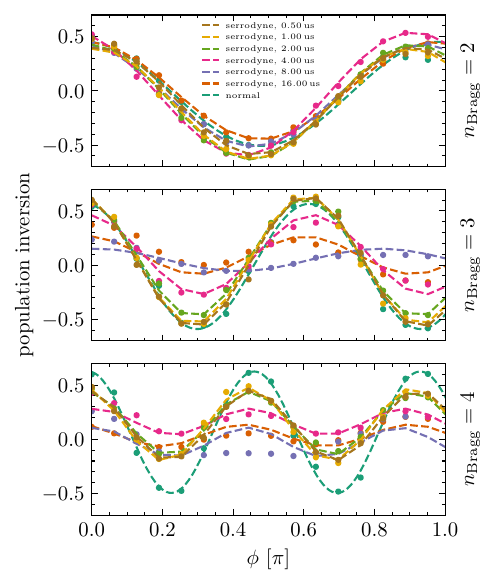}
    \caption{MZI fringes recorded with normal and serrodyne-modulated Bragg pulses for different Bragg orders. Dots are experimentally recorded data points. Dashed lines are fits to those points.}
    \label{fig:mzi_fringes}
\end{figure}

The population in the $\ket{0}$ output port is given by $\cos^2((4nT^2kg-\nbragg\phi)/2)$, and the population in the $\ket{\nbragg}$ output port is given by $\sin^2((4nT^2kg-\nbragg\phi)/2)$. Because of the gravity ramp described in \fig{ifr_chain_diagram}, $g=0$ for the MZI implemented here. From these population equations we see that the fringe frequency will vary as a function of $\nbragg$ as $\phi$ is scanned. This is observed experimentally as shown in \fig{mzi_fringes}. These fringes also reveal that non-zero fall durations can lead to contrast reductions and interferometer phase shifts. The sawtooth waveform phase used for each MZI was selected using the same method described in section \ref{sec:serrodyne_single_bragg}.

\begin{figure}
    \centering
    \includegraphics[width=0.5\linewidth]{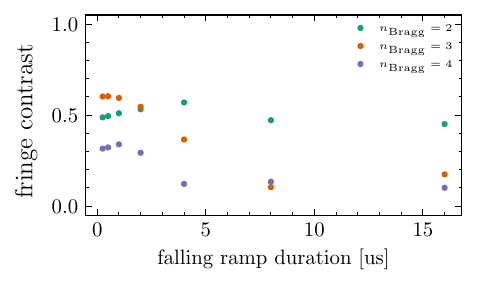}
    \caption{Measured MZI fringe contrast as a function of the falling ramp duration and the ramp amplitude for different Bragg orders. Contrasts are measured by fitting a sinusoid to MZI fringes as in \fig{mzi_fringes} and extracting the amplitude. The theoretical maximum contrast of the interferometer is $1$, and is limited to a maximum of approximately $0.6$ due to the variation of $\oeff$ across the atom cloud.}
    \label{fig:mzi_contrast_scan}
\end{figure}

In \fig{mzi_contrast_scan} we show the dependence of the MZI fringe contrast on the falling ramp duration and ramp amplitude. We observe that lower Bragg orders generally achieve a higher contrast at long fall durations, which can be partially attributed to the smaller fraction of the sawtooth period that these fall durations account for. 

\section{Conclusion}

We have proposed an interferometer design that both simplifies laser requirements and breaks the degeneracy produced by directional symmetry in retroreflected atom interferometers by serrodyne-modulating the retroreflection mirror. We experimentally verified this approach in an atomic fountain and tested the performance of the Bragg diffraction pulses for a range of sawtooth-waveforms. We found that single Bragg pulses are viable for suboptimal sawtooth waveform parameters, and verified that parameters still produce working interferometers. The observation of viable single Bragg pulses and full interferometers for poor quality serrodyne-modulation implies that the Bragg diffraction process is limited by the serrodyne-modulated Bragg diffraction process and not by the purity of the serrodyne-modulated frequency. We expect that this method will find use in the development of atom interferometers requiring optical simplicity, or those needing a method by which to break the degeneracy produced by directional symmetry.

\appendix*
\section{Derivation of the equations of motion}
\label{sec:derivation_eom}

We begin with the equations of motion for a two-level system in an arbitrary number of laser fields
\begin{align}
    \dot{c}_e(z,t)=&-i\sum_\ell\Omega_\ell\cos(k_\ell z-\omega_\ell t+\theta_\ell)c_g(z,t) \\
    \dot{c}_g(z,t)=&-i\omega_{eg}c_g(z,t)-i\sum_\ell\Omega_\ell\cos(k_\ell z-\omega_\ell t+\theta_\ell)c_e(z,t)
\end{align}
We transform into a frame rotating at $\omega_{eg}$ and perform the rotating wave approximation to find
\begin{align}
    \dot{c}_e(z,t)=&-i\sum_\ell\frac{\Omega_\ell}{2}\exp(i(-k_\ell z-\Delta_\ell t-\theta_\ell))c_g(z,t) \\
    \dot{c}_g(z,t)=&-i\sum_\ell\frac{\Omega_\ell}{2}\exp(i(k_\ell z+\Delta_\ell t+\theta_\ell))c_e(z,t)
\end{align}
where $\Delta_\ell=\omega_{eg}-\omega_\ell$. We redefine $\Omega_\ell\exp(i(-k_\ell z-\theta_\ell))\to\Omega_\ell$, finding
\begin{align}
    \dot{c}_e(z,t)=&-i\sum_\ell\frac{\Omega_\ell}{2}\exp(-i\Delta_\ell t)c_g(z,t) \label{eq:eom_ce} \\
    \dot{c}_g(z,t)=&-i\sum_\ell\frac{\Omega_\ell^*}{2}\exp(i\Delta_\ell t)c_e(z,t)
\end{align}
We then adiabatically eliminate the excited state (assuming all fields are far detuned from the atomic transition) by integrating \eq{eom_ce} and pulling $c_g(z,t)$ and $\Omega_\ell$ out of the integral, finding
\begin{eqnarray}
    c_e(z,t)\approx\sum_\ell\frac{\Omega_\ell}{2\Delta_\ell}\exp(-i\Delta_\ell t)c_g(z,t)
\end{eqnarray}
Note that $\Omega_\ell$ can be time-dependent if the field phase or amplitude varies in time. Plugging this into our expression for $\dot{c}_g(z,t)$, expanding back out the Rabi frequencies, and expanding out the definition for $\Delta_\ell$ in the exponential yields
\begin{widetext}
\begin{eqnarray}
    \dot{c}_g(z,t)=-i\sum_{\ell,\ell'}\frac{\Omega_{\ell'}\Omega_\ell}{4\Delta_\ell}\exp(i[(k_{\ell'}-k_\ell)z-(\omega_{\ell'}-\omega_\ell)t+\theta_{\ell'}-\theta_\ell])c_g(z,t)
\end{eqnarray}
\end{widetext}
We consider a four field system with the following parameter values
\begin{gather}
    \omega_3=\omega_1\qquad\omega_4=\omega_2 \nonumber \\
    k_1=k_2=k\qquad k_3=k_4=-k \\
    \Omega_3=\Omega_1\qquad \Omega_4=\Omega_2 \nonumber
\end{gather}
We assume that $\omega_1$ and $\omega_2$ are order $1\,\mathrm{kHz}$ to $1\,\mathrm{MHz}$ apart. We also assume both of these are optical frequencies (i.e. several hundred $\mathrm{THz}$). This means that we can approximate each $1/\Delta_\ell$ in the sum as the same $1/\Delta$. Expanding out the sum and defining $\delta=\omega_1-\omega_2$ yields
\begin{widetext}
\begin{eqnarray}
    \dot{c}_g(z,t)=&-\frac{i}{2\Delta}\left[\Omega_1^2+\Omega_2^2+\Omega_1^2\cos(\theta_1-\theta_3+2kz)+\Omega_2^2\cos(\theta_2-\theta_4+2kz)\right. \nonumber \\
    &+\Omega_2\Omega_1\cos(\theta_2-\theta_3+2kz+\delta t)+\Omega_2\Omega_1\cos(\theta_1-\theta_4+2kz-\delta t) \nonumber \\
    &\left.+\Omega_2\Omega_1\cos(-\theta_1+\theta_2+\delta t)+\Omega_2\Omega_1\cos(-\theta_3+\theta_4+\delta t)\right]c_g(z,t)
\end{eqnarray}
\end{widetext}
We add a kinetic energy term $\hat{p}^2c_g(z,t)/(2m)$, transform to momentum space, and define $c_{n}(t)=c_g(2n\hbar k,t)$ to find
\begin{widetext}
\begin{align}
    \dot{c}_n(t)=&-i4n^2\omegar c_n(t) \nonumber \\
    &-\frac{i}{4\Delta}\qty[2\Omega_1^2+2\Omega_2^2+2\Omega_1\Omega_2\cos(t\delta-\theta_1-\theta_2)+2\Omega_1\Omega_2\cos(t\delta-\theta_3-\theta_4)]c_n(t) \nonumber \\
    &-\frac{i}{4\Delta}\bigl[\Omega_1^2\exp(i(\theta_1-\theta_3))+\Omega_2^2\exp(i(\theta_2-\theta_4)) \nonumber \\
    &\phantom{-\frac{i}{4\Delta}\bigl[}+\Omega_1\Omega_2\exp(i(\delta t+\theta_2-\theta_3))+\Omega_1\Omega_2\exp(i(-\delta t+\theta_1-\theta_4))\bigl]c_{n-1}(t) \nonumber \\
    &-\frac{i}{4\Delta}\bigl[\Omega_1^2\exp(-i(\theta_1-\theta_3))+\Omega_2^2\exp(-i(\theta_2-\theta_4)) \nonumber \\
    &\phantom{-\frac{i}{4\Delta}\bigl[}+\Omega_1\Omega_2\exp(-i(\delta t+\theta_2-\theta_3))+\Omega_1\Omega_2\exp(-i(-\delta t+\theta_1-\theta_4))\bigl]c_{n+1}(t) \label{eq:eom_cg_momentum}
\end{align}
\end{widetext}
We now consider the two specific cases relevant for this paper. The first is the case of a moving mirror with a single laser and its retroreflection. In this system $\Omega_2=\Omega_4=0$, $\Omega_1=\Omega_3=2\sqrt{\oeff\Delta}$, $\omega_1=\omega_3$, $k_1=-k_3=k$, $\theta_1=0$, and $\theta_3=\theta(t)=2kz(t)-\pi$. Applied to \eq{eom_cg_momentum} we find
\begin{eqnarray}
    \dot{c}_n(t)=&-i4n^2\omegar c_n(t)-i2\oeff c_n(t) \nonumber \\
    &-i\oeff\exp(-i\theta(t))c_{n-1}(t) \nonumber \\
    &-i\oeff\exp(i\theta(t))c_{n+1}(t) \label{eq:eom_mirror}
\end{eqnarray}
The second case is the case of a stationary mirror with a two-frequency laser and its retroreflection. In this system $\Omega_1=\Omega_2=\Omega_3=\Omega_4=2\sqrt{\oeff\Delta}$, $k_1=k_2=-k_3=-k_4=k$, $\omega_3=\omega_1$, $\omega_4=\omega_2$, $\theta_2=\theta_4=0$, and $\theta_1=\theta_3+\pi=-\theta(t)$. Applied to \eq{eom_cg_momentum} we find
\begin{eqnarray}
    \dot{c}_n(t)=&-i4n^2\omegar c_n(t)-i4\oeff c_n(t) \nonumber \\
    &-2\oeff\sin(\delta t+\theta(t))c_{n-1}(t) \nonumber \\
    &+2\oeff\sin(\delta t+\theta(t))c_{n+1}(t)
\end{eqnarray}
In an atomic fountain the wavefunction is initially in a high $n$ state, which allows us to make the rotating wave approximation. To do this we perform the transformation $c_n(t)\to c_n(t)\exp(-i4n^2\omegar t)$, yielding
\begin{widetext}
\begin{eqnarray}
    \dot{c}_n(t)=-i4\oeff c_n(t)-2\oeff\sin(\delta t+\theta(t))\exp(-i4\omegar t)\qty[\exp(i8n\omegar t)c_{n-1}(t)-\exp(-i8n\omegar t)c_{n+1}(t)]
\end{eqnarray}
\end{widetext}
If the sine term is expanded into exponentials and we choose $\delta$ and $n$ to be positive, then we can make the rotating wave approximation to find
\begin{align}
    \dot{c}_n(t)=&-i4\oeff c_n(t) \nonumber \\
    &-i\oeff\exp(i(-\delta t-\theta(t)-4\omegar t+8n\omegar t))c_{n-1}(t) \nonumber \\
    &-i\oeff\exp(i(\delta t+\theta(t)-4\omegar t-8n\omegar t))c_{n+1}(t)
\end{align}
Reverting the frame transformation leads to
\begin{eqnarray}
    \dot{c}_n(t)=&-i(4\oeff+4n^2\omegar-n\delta)c_n(t) \nonumber \\
    &-i\oeff\exp(-i\theta(t))c_{n-1}(t) \nonumber \\
    &-i\oeff\exp(i\theta(t))c_{n+1}(t)
\end{eqnarray}
In our atomic fountain experiment we simulate these equations of motion by retroreflecting a two-frequency far-detuned laser off of a stationary mirror with an atom at a high initial $n$ value. If $\delta=4n\omegar$ then we see this result is equivalent to \eq{eom_mirror} up to a factor of 2 in the AC Stark shift term. In both equations the AC Stark shift term can be dropped entirely as it contributes a common phase to both interferometer arms, and therefore does not contribute to the interferometer phase.

\begin{acknowledgments}
Safran acknowledges funding under the DARPA Robust Quantum Sensing (RoQS) program under DARPA Contract No. HR0011-25-9-0207. The views, opinions, and/or findings expressed are those of the authors and should not be interpreted as representing the official views or policies of DARPA or the U.S. Government.
\end{acknowledgments}

\bibliography{references}

\end{document}